\documentclass[a4paper,11pt]{article}
\usepackage{jcappub} % for details on the use of the package, please
\usepackage[utf8]{inputenc}
\usepackage[T1]{fontenc}

\usepackage[mathscr]{eucal}
\usepackage[Symbolsmallscale]{upgreek}
\usepackage{xcolor}
\usepackage{slashed}
\usepackage{hyperref}
\usepackage{soul}

\usepackage{bm,epstopdf,natbib,hyperref,color,verbatim,multirow,bm,tikz,xcolor}
\hypersetup{colorlinks=true,urlcolor=blue,citecolor=blue,linkcolor=blue,menucolor=blue,anchorcolor=blue,filecolor=blue}
\everymath{\displaystyle}
\definecolor{lime}{HTML}{A6CE39}
\DeclareRobustCommand{\orcidicon}{
	\begin{tikzpicture}
	\draw[lime, fill=lime] (0,0) 
	circle [radius=0.16] 
	node[white] {{\fontfamily{qag}\selectfont \tiny ID}};
	\draw[white, fill=white] (-0.0625,0.095) 
	circle [radius=0.007];
	\end{tikzpicture}
	\hspace{-3mm}
}
\foreach \x in {A, ..., Z}{\expandafter\xdef\csname orcid\x\endcsname{\noexpand\href{https://orcid.org/\csname orcidauthor\x\endcsname}
			{\noexpand\orcidicon}}
}

\title{
Thermodynamic Consistent Description of Compact Stars of Two Interacting Fluids: The Case of
Neutron Stars with Higgs Portal Dark Matter 
}

\author[a]{Fazlollah Hajkarim\orcidA{},%\note{Corresponding author.}
}
\author[b]{J{\"u}rgen Schaffner-Bielich\orcidB{},}

\author[c,d,e]{Laura Tolos\orcidC{},}

\affiliation[a]{Department of Physics and Astronomy, University of Oklahoma, Norman, OK 73019, USA}
\affiliation[b]{Institut f\"ur Theoretische Physik,
Goethe Universit\"at Frankfurt, Max-von-Laue-Str. 1,\\ D-60438 Frankfurt am Main, Germany}
\affiliation[c]{Institute of Space Sciences, Campus UAB, \\ Carrer de Can Magrans, 08193 Barcelona, Spain}
\affiliation[d]{Institut d’Estudis Espacials de Catalunya, 08034 Barcelona, Spain}
\affiliation[e]{Frankfurt Institute for Advanced Studies, \\ Ruth-Moufang-Str. 1, 60438 Frankfurt am Main, Germany}

\emailAdd{fazlollah.hajkarim@ou.edu}
\emailAdd{schaffner@astro.uni-frankfurt.de}
\emailAdd{tolos@ice.csic.es}

\abstract{
We consider a thermodynamically consistent approach for the computation of the masses, radii, and tidal deformabilities of 
compact stars consisting of two interacting fluids with separately conserved quantum numbers. 
We apply this interacting fluid approach to the case of compact stars of neutron star matter with 
the Higgs portal fermionic dark matter model for the first time in a thermodynamically consistent manner. 
The patterns for the mass-radius curves and the tidal deformability depend on the dark matter particle mass and are different from former studies.
 Compared to ordinary neutron star properties, we obtain smaller masses and radii for dark matter particle masses similar to the nucleon mass and, hence, smaller tidal deformabilities 
as a result of the softening of the equation of state due to the presence of dark matter. For dark matter particle masses below the nucleon mass and sizable chemical potentials with respect to the dark matter particle mass, there will be a dark halo instead of dark core. 
Our investigation provides the basis for 
studying mergers of compact stars where the two fluids of neutron star matter and dark matter are coupled kinetically to each other and are described by one combined energy-momentum tensor of the two interacting fluids but are chemically different with two separately conserved number currents. 
}

\begin{document}
\maketitle
\flushbottom

\section{Introduction}
\label{sec:intro}
Dark matter (DM), an enigmatic and invisible component of the universe, constitutes about 85\% of the total matter content~\cite{Cirelli:2024ssz}. Its existence would be primarily determined through gravitational effects on visible matter, radiation, and the large-scale structure of the cosmos. Despite extensive searches, the precise nature and composition of DM remain unresolved.  

Among others, DM could  
influence the physical properties and dynamics of compact stars, such as neutron stars, thus leading to a deeper understanding of these objects and the broader dark sector \cite{Antoniadis:2013pzd,Bauswein:2020kor,Berezhiani:2020zck,Bose:2021yhz,
Bramante:2023djs,Das:2018frc,DelPopolo:2020hel,Demorest:2010bx,Diedrichs:2023trk,
Ellis:2017jgp,Ellis:2018bkr,Giangrandi:2022wht,Hossain:2020mvn,
Hossain:2021qyg,Karkevandi:2021ygv,Shakeri:2022dwg,Kouvaris:2011gb,Li:2022url,Maselli:2019ubs,
Miao:2022rqj,Narain:2006kx,Nelson:2018xtr,Profumo:2006bv,Rahimi:2024qjl,Raj:2017wrv,
Ryan:2022hku,Sagun:2021oml,Thakur:2023aqm,Thakur:2024btu,Thakur:2024ejl,
Thakur:2024mxs,Tolos:2015qra,Tolos:2017lgv,Tolos:2016hhl,Vikiaris:2023vau, DelGrosso:2024wmy,Mariani:2023wtv}. Indirect detection possibilities, such as neutrino and gamma-ray signals from inelastic DM annihilations near neutron stars, further increase the potential for uncovering DM properties \cite{Acevedo:2024ttq}.
Neutron stars can play the role of DM detectors and provide more information about DM properties using the study of the DM interactions with nucleons~\cite{Raj:2017wrv, 2104.02700,Vikiaris:2023vau,Ray:2023auh,Cline:2018ami,Husain:2022brl}.
Furthermore, the concept of supermassive dark objects broadens insights into DM's influence on stellar evolution, and white dwarfs mixed with DM may collapse into neutron stars, affecting the overall stellar population \cite{Vikiaris:2023vau, Chang:2018bgx, Leung:2019ctw, Gresham:2018rqo}. 

Gravitational wave astronomy is another possible tool to detect DM  within neutron stars. Observations from experiments such as LIGO (Laser Interferometer Gravitational-Wave Observatory) and Virgo, including events like GW170817, provide insights into the structure of neutron stars \cite{LIGOScientific:2017vwq,LIGOScientific:2018cki}. Non-radial oscillation modes in DM-admixed neutron stars suggest that these oscillations could serve as indicators of DM presence \cite{Shirke:2023ktu, Thakur:2024ejl, Shirke:2024ymc, Thakur:2023aqm,Kumar:2024abb}. Additionally, the study of tidal Love numbers--parameters that quantify a star's tidal response to external gravitational fields--enhances our understanding of the composition of neutron stars \cite{Hinderer:2007mb, Hinderer:2009ca, Postnikov:2010yn, Zacchi:2020dxl} and allows for elucidating the possible existence and nature of DM.

Specific DM candidates, including axions, dark photons, and ultralight particles, exhibit unique astrophysical signatures that can be probed through neutron star studies. Axions, for example, may address the cusp-core problem in galactic halos and significantly influence neutron star cooling processes \cite{Marsh:2015wka, Sedrakian:2015krq, Sedrakian:2018kdm}. Dark photons impact neutron star stability and could generate observable signals detectable through various astrophysical observations \cite{Flambaum:2019cih, Alexander:2020wpm, Cline:2018ami, Chang:2018bgx}. Ultralight dark photons and millicharged particles extend DM models with significant implications for neutron star stability and dynamics \cite{Flambaum:2019cih, Alexander:2020wpm, Cline:2022leq}, while studies suggesting neutron decay into a dark Dirac fermion and a dark photon could explain discrepancies in neutron lifetime measurements, offering new pathways for DM detection \cite{Cline:2018ami, Husain:2022brl}.  Neutron stars can also be used for the detection of DM and probe DM properties through their interactions  with nucleons within the core of neutron stars~\cite{Ray:2023auh,Cline:2018ami,Husain:2022brl}. Exotic compact objects, such as mirror neutron stars and boson stars, offer valuable insights into DM's astrophysical role. Mirror matter within neutron stars could produce distinctive gravitational wave signatures, differentiating these stars from their ordinary counterparts \cite{Hippert:2021fch, Ciancarella:2020msu, Berezhiani:2020zck}. 
 Self-interacting DM models, particularly those involving the Higgs portal and muonic forces, are investigated using astrophysical data to verify their viability \cite{Kouvaris:2014uoa, Dror:2019uea,Garani:2019fpa}.
Boson stars, formed from bosonic DM with repulsive self-interactions, present alternative compact objects that challenge traditional neutron star models \cite{Agnihotri:2008zf, DiGiovanni:2021ejn, Collier:2022cpr,Visinelli:2021uve,Guo:2019sns,Pitz:2023ejc}. Studies on fermion-boson stars aim to reconcile observations of compact stars with complex DM interactions, bridging theoretical predictions with astronomical data \cite{Diedrichs:2023trk,DiGiovanni:2021ejn, Alexander:2016glq, Cassing:2022tnn,Pitz:2024xvh}. 
Moreover, asymmetric bosonic DM, mediated by vector fields coupled to scalar particles, may alter neutron stars' mass, radius, and tidal deformability \cite{Giangrandi:2022wht, Panotopoulos:2017idn, Bhat:2019tnz,Karkevandi:2021ygv,Shakeri:2022dwg,Barbat:2024yvi,Dengler:2021qcq}. Non-annihilating particle DM can slowly accumulate within neutron stars, potentially leading to the formation of low-mass black holes, a scenario supported by LIGO-Virgo observations \cite{Ray:2023auh,Gresham:2018rqo,Leung:2019ctw,Kouvaris:2015rea,Fuller:2014rza,Bauswein:2020kor}. DM may also influence neutron star heating and cooling, with density-dependent EoS models showing that DM admixtures impact thermal evolution and axion emissions from protoneutron stars suggest possible indirect detection of these  particles \cite{Bhat:2019tnz, Sedrakian:2015krq, Sedrakian:2018kdm, Yakovlev:2004iq, Hutauruk:2023nqc, Fischer:2021jfm,Bar:2019ifz, Zeng:2021moz,Gau:2023rct,Fortin:2018aom,Fortin:2018ehg}. These cooling mechanisms, governed by energy loss processes such as neutrino emission, can constrain axion properties and their coupling strengths to ordinary matter, offering a window into physics beyond the Standard Model \cite{Yakovlev:2004iq, Sedrakian:2015krq, Sedrakian:2018kdm}.
 
In this work we investigate the Higgs portal fermionic DM model, where DM interacts with nucleonic matter through a Higgs mediator, enabling the coupling to nucleons. In order to explore the mass-radius relationship of compact astrophysical objects composed of both dark and visible matter, we solve the structure equations for the combined system. Several approaches treat dark and visible matter as interacting fluids typically assuming fixed Fermi momenta for DM~
\cite{Lenzi:2022ypb, Thakur:2024btu, Sen:2024yim, Pal:2024afl, Kumar:2024zzl, Quddus:2019ghy, Das:2018frc, Panotopoulos:2017idn, Lopes:2024ixl}. This procedure leads to thermodynamic inconsistencies, as the Fermi momentum and, hence, the chemical potential are expected to vary with radial distance within the object. In contrast, we pursue a thermodynamically consistent treatment of interacting dark and visible matter by solving the structure equations within the interacting two-fluid approach (IFA) with separately conserved quantum numbers and associated chemical potentials following the description of Kodama and Yamada \cite{kodama1972theory}. We stress that IFA is different to the canonical two-fluid approach where the two fluids are not interacting with each other so that there are two separately conserved energy-momentum tensors. The IFA applies to all cases where two fluids are coupled such that only one energy-momentum tensor for both fluids is conserved, so there is kinetic coupling, but are chemical different, so that there are two conserved number currents for the two fluids. 

The paper is organized as follows. In Sec.~\ref{sec:tov} we discuss the structure equations and the different approaches to solve them, pointing out the necessity of a thermodynamically consistent treatment of the interacting nucleonic and DM system. Then, in Sec.~\ref{sec:meanfield}  we use the mean field model for nucleons  and Higgs portal DM to determine the equation of state (EoS) of IFA. Using this EoS, we compute the structure equations and find the mass-radius relation in Sec.~\ref{sec:massrad} for different DM particle masses.  Moreover,  we compute the tidal deformability and compare our outcome with the  tidal deformability constraint  from LIGO  observation in Section~\ref{sec:tidal}. Finally, we present our summary and conclusions in Sec.~\ref{sec:conc}.

\section{Structure Equations: Two Approaches}
\label{sec:tov}

Thermodynamic consistency requires expanding the equilibrium conditions for relativistic fluids in gravitational fields. Klein’s law generalizes classical thermodynamic equilibrium to account for variations in the chemical potential $ \mu $ for gravitating systems, showing that equilibrium requires \cite{klein1949thermodynamical,Haensel:2007yy,kodama1972theory,Glendenning:1997wn}
\begin{eqnarray}
\label{eq:klein}
\mu e^{\nu(r)} = \text{constant},
\end{eqnarray}
where $ g_{00} = e^{2\nu(r)} $ is the time-time  component of the gravitational metric. Klein’s framework shows that self-gravitating fluids reach equilibrium when the chemical potential appropriately adjusts according to the gravitational potential. This law complements the Tolman-Ehrenfest (TE) law \cite{Tolman:1930zza, 1930PhRv...36.1791T}, which describes temperature gradients in a gravitational field for thermal equilibrium. 
From Eq.~(\ref{eq:klein}) we can obtain the radial dependency of chemical potential $ \mu(r) $ of one-fluid in a gravitational field
is given by~\cite{Haensel:2007yy,kodama1972theory,Glendenning:1997wn}
\begin{eqnarray}
\label{eq:mudiff}
\mu(r) = \mu_{c} e^{-\nu(r)}, 
\end{eqnarray}
where $ \mu_{c} $ is the chemical potential at the star’s center. This expression emerges from the hydrostatic equilibrium condition in relativistic systems, ensuring that the chemical potential varies in response to the gravitational redshift.

By differentiating both sides of Eq.~(\ref{eq:mudiff}) with respect to the radial coordinate $ r $, we can eliminate the constant chemical potential $ \mu_{c} $
\cite{Haensel:2007yy,kodama1972theory,Glendenning:1997wn}%
\begin{eqnarray}
\label{eq:mu-diff}
\frac{d\mu(r)}{dr} = -\mu(r) \frac{d\nu(r)}{dr}\,.
\end{eqnarray}
This differential equation relates the spatial variation of the chemical potential directly to the gradient of the metric function $ \nu(r) $, which encapsulates the gravitational field inside the star.

The radial coordinate $ r $ appears explicitly in the expression for the derivative of the metric function~\cite{Glendenning:1997wn,Shapiro:1983du}
\begin{eqnarray}
\label{eq:nu-diff}
\frac{d\nu(r)}{dr} = \frac{  G \left(M(r) + 4\pi r^3 p(r) \right)}{r \left( r - 2 G M(r) \right)}\,,
\end{eqnarray}
where $ M(r) $ is the mass enclosed within radius $ r $, $ p(r) $ is the pressure and $ G $ is Newton  gravitational constant. Note that $c=1$. This equation is derived from the Einstein field equations under the assumption of a static, spherically symmetric mass distribution.

Equivalently, one can formulate the equilibrium conditions for one-fluid in terms of the standard structure equations, or Tolman-Oppenheimer-Volkov (TOV) equations~\cite{Tolman:1939jz,Oppenheimer:1939ne}
\begin{eqnarray}
\frac{dp(r)}{dr} &=& -\left( \epsilon(r) + p(r) \right) \frac{d\nu(r)}{dr}\,, \label{eq:TOV_pressure}
\end{eqnarray}
with
\begin{eqnarray}
\frac{dM(r)}{dr} &=& 4\pi r^2 \epsilon(r)\, \label{eq:TOV_mass}  ,
\end{eqnarray}
where $ \epsilon(r) $ is the energy density. Equation~\eqref{eq:TOV_mass} describes how mass accumulates with radius due to the energy density, while Eq.~\eqref{eq:TOV_pressure} ensures that the pressure gradient balances the gravitational pull. 

This equivalence results  from connecting the chemical potential with the energy density $ \epsilon $, pressure density $ p $, and number density $ \rho $ as~\cite{Haensel:2007yy} 
\begin{eqnarray}
\mu = \frac{\epsilon + p}{\rho}\,,
\end{eqnarray}
or, alternatively, it can be derived from the energy density as the derivative with respect to the number density \footnote{
For nucleonic matter, one can solve the structure equations based on the baryon chemical potential $ \mu_{\rm N} $ by considering the baryon/nucleon number density $ \rho_{\rm N} $, which includes both neutrons and protons
\begin{eqnarray}
\mu_{\rm N} = \frac{\epsilon + p}{\rho_{\rm N}}\,, \quad \mu_{\rm N} = \mu_n + \mu_p\,, \quad \rho_{\rm N} = \rho_n + \rho_p\,.
\end{eqnarray}
This approach allows for a unified treatment of the matter inside neutron stars, accounting for the contributions from different particle species.}
\begin{eqnarray}
\mu = \frac{d\epsilon}{d \rho}\,.
\end{eqnarray}

In modeling self-gravitating systems, particularly neutron stars with DM components, it is important to distinguish between one-fluid and two-fluid cases. The conservation of the energy-momentum tensor provides a foundation for understanding these scenarios.
When the visible and dark matter components are considered as a unified fluid, the conservation of the total energy-momentum tensor ensures hydrostatic equilibrium. This combined system adheres to a single structure equation, where both pressure and chemical potential gradients are balanced against gravitational forces. The energy-momentum tensor $ T^{\mu\nu} $ satisfies \cite{Glendenning:1997wn,Shapiro:1983du}
   \begin{eqnarray}
       \nabla_{\mu} T^{\mu\nu} = 0\,,
   \end{eqnarray}
   ensuring equilibrium across the entire mixture, with mutual interactions effectively described through a unified set of thermodynamic variables.
   
When the visible and dark matter are treated as distinct interacting fluids, each with its own energy-momentum tensor $ T^{\mu\nu}_{\rm N} $ and $ T^{\mu\nu}_{\rm DM} $, the total energy-momentum conservation still holds
   \begin{eqnarray}
   \nabla_{\mu} (T^{\mu\nu}_{\rm N} + T^{\mu\nu}_{\rm DM}) = 0\,.
   \end{eqnarray}
   However, internal interactions allow for energy-momentum exchange between the fluids, described by
   \begin{eqnarray}
   \nabla_{\mu} T^{\mu\nu}_{\rm N} = -Q^{\nu}, \quad \nabla_{\mu} T^{\mu\nu}_{\rm DM} = Q^{\nu}\,,
   \end{eqnarray}
   where $ Q^{\nu} $ represents the interaction term. This formulation accommodates scenarios where dark matter impacts the visible matter’s thermodynamics, altering temperature and chemical potential distributions and modifying the resulting gravitational equilibrium.
Understanding these cases is crucial for accurately modeling the structure and behavior of compact astrophysical objects under extreme gravitational conditions.

When considering the case of DM interacting with ordinary matter, we have two interacting but separately conserved matter species. Then, in order to solve the structure of the DM admixed-compact objects it is necessary to generalize Eqs.~(\ref{eq:mu-diff},\ref{eq:nu-diff}) to the two-fluid interacting system by conserving each chemical potential $\mu_i$ separately \cite{kodama1972theory,Gresham:2018rqo}. As a result, the structure equations read
\begin{eqnarray}
&&\frac{d\mu_i(r)}{dr} = -\mu_i(r) \frac{d\nu(r)}{dr}\, ,
\label{eq:anagTOV1}\\
&&\frac{d\nu(r)}{dr} = \frac{  G \left(M(r) + 4\pi r^3 p(r) \right)}{r \left( r - 2 G M(r) \right)}\, ,
\label{eq:anagTOV2} \\
&&\frac{dM(r)}{dr} = 4\pi r^2 \epsilon(r)\, \label{eq:TOV_mass2} ,
\end{eqnarray}
with the total mass of the system given by $M=\sum_i M_i$, with $M_i$ the mass contributions of DM and ordinary matter. The total pressure is obtained from $p=\sum_i \partial \epsilon/ \partial \rho_i \ \rho_i - \epsilon $, with $\rho_i$ the individual contribution to the number density of each species and $\epsilon$ the total energy density of the interacting system.

When the interaction DM-nucleons is absent, a given species affects the other only through the total gravitational mass and pressure. Then, only in this  case, one recovers the standard formulation of the TOV equations for two fluids with separately conserved energy-momentum tensors, see the discussion in the appendix of Ref.~\cite{Gresham:2018rqo}: 
\begin{eqnarray}
&&\frac{dp_i(r)}{dr} = -(\epsilon_i(r)+p_i(r)) \frac{d\nu(r)}{dr}\, ,
\label{eq:TOV1}\\
&&\frac{d\nu(r)}{dr} = \frac{  G \left(M(r) + 4\pi r^3 p(r) \right)}{r \left( r - 2 G M(r) \right)}\, .
\label{eq:TOV2} \\
&&\frac{dM(r)}{dr} = 4\pi r^2 \epsilon(r)\, , \label{eq:TOV_mass3}
\end{eqnarray}
where $\epsilon_i$ and $p_i$ stand for the energy density and pressure of each species, respectively, with $\epsilon=\sum_i \epsilon_i$ and $p=\sum_i p_i$.

Lately, several approaches treat DM and ordinary matter as two-interacting fluids but solve the standard TOV equations by assuming fixed Fermi momenta for DM \cite{Lenzi:2022ypb, Thakur:2024btu, Sen:2024yim, Pal:2024afl, Kumar:2024zzl, Quddus:2019ghy, Das:2018frc, Panotopoulos:2017idn, Lopes:2024ixl}. This procedure leads to thermodynamic inconsistencies, as the Fermi momentum and, hence, the chemical potential would vary within the star, as determined by Eqs.~(\ref{eq:anagTOV1},\ref{eq:anagTOV2},\ref{eq:TOV_mass2}). 

\subsection{Dimensionless structure equations for the interacting DM-nucleonic system}

To facilitate the solution of Eqs.~(\ref{eq:anagTOV1},\ref{eq:anagTOV2}), it is common to redefine variables using dimensionless quantities. The gravitational constant $ G $ can be expressed in terms of the Planck mass $ M_{\rm Pl} $ as $ G \equiv M_{\rm Pl}^{-2} $. Introducing the scaling factors (Landau mass $M_L$ and radius $R_L$) ~\cite{Shapiro:1983du,Glendenning:1997wn,Narain:2006kx}
\begin{eqnarray}
M' = \frac{M}{M_{\rm L}}\,, \quad M_{\rm L} = \frac{M_{\rm Pl}^3}{m_{\rm f}^2}\,, \quad r' = \frac{r}{R_{\rm L}}\,, \quad R_{\rm L} = \frac{M_{\rm Pl}}{m_{\rm f}^2}\,, \quad\mu' = \frac{\mu}{m_f}\,,
\end{eqnarray}
where the fermion mass $ m_{\rm f} $ is typically taken as the nucleon mass $ M_{\rm N} = (M_p + M_n)/2 $. These scaling factors normalize the mass and radius in terms of fundamental constants, simplifying the numerical integration.

Using these scaled variables, the structure equations can be rewritten in dimensionless form
 in terms of the metric function, chemical potentials and the total mass of compact object ~\cite{Glendenning:1997wn,Narain:2006kx}
\begin{eqnarray}
\frac{d\nu(r')}{dr'} &=& \frac{   \left( M'(r') + 4\pi r'^3 p'(r') \right)}{r' \left( r' - 2  M'(r') \right)}\,, \label{eq:chem1} \\
\frac{dM'(r')}{dr'} &=& 4 \pi r'^{2} \epsilon'(r') \,, \label{eq:chem2} \\
\frac{d\mu'_{\rm N}(r')}{dr'} &=& -\mu'_{\rm N}(r') \frac{d\nu(r')}{dr'}\,, \label{eq:chem3}  \\
\frac{d\mu'_{\rm DM}(r')}{dr'} &=& -\mu'_{\rm DM}(r') \frac{d\nu(r')}{dr'}\,.
\label{eq:chem4}
\end{eqnarray}
Here, $p' = p / m_f^4 $ and $\epsilon' = \epsilon / m_f^4 $ are the dimensionless pressure and energy density, respectively. These equations are now suitable for numerical integration starting from the center of the star, where $ r' = 0 $,  $ \nu'(0) = 0 $, $ M'(0) = 0 $, $ \epsilon'(0)=\epsilon'_{c} $, and $ p'(0) = p'_{c} $, with  $\mu'_{{\rm N},c}$ and  $\mu'_{{\rm DM},c}$ being the fixed central chemical potentials for nucleons and DM, respectively. 
 The integration of the structure equation for the nucleon chemical potential finishes when the nucleon chemical potential reaches the nucleon mass. The same happens for the DM component, with the calculation ending when the DM chemical potential equals the DM particle mass.

\section{The Higgs Portal Fermionic Dark Matter Model}
\label{sec:meanfield}

 Relativistic mean-field (RMF) theory has been a powerful tool in describing the properties of hadronic matter under extreme conditions~\cite{Serot:1984ey}. Recently, the potential impact of  DM on neutron star properties has attracted  significant interest~\cite{Bertone:2007ae,Cardoso:2019rvt,Tolos:2015qra}. In this work, we make use of the  nucleonic RMF model and extend it by including the interactions between nucleons and DM particles via the Higgs portal, aiming to investigate the effects of such interactions on the EoS of neutron star matter.

In the RMF framework, the nucleon-meson interaction is described by the Lagrangian density~\cite{Das:2018frc,Das:2021hnk,Serot:1984ey, Walecka:1974qa}
\begin{align}
\mathcal{L} =&\ \bar{\psi} \left[ \gamma^\mu \left( i\partial_\mu - g_v V_\mu - g_\rho \tau \cdot b_\mu \right) - \left( M_{\rm N} + g_s \phi \right) \right] \psi \nonumber \\
&+ \frac{1}{2} \partial_\mu \phi \partial^\mu \phi - \frac{1}{2} m_s^2 \phi^2 - \frac{1}{3} g_2 \phi^3 - \frac{1}{4} g_3 \phi^4 \nonumber \\
&- \frac{1}{4} V_{\mu \nu} V^{\mu \nu} + \frac{1}{2} m_V^2 V_\mu V^\mu - \frac{1}{4} b_{\mu \nu} b^{\mu \nu} + \frac{1}{2} m_\rho^2 b_\mu b^\mu\,,
\label{eq:lagrangian_nucleon}
\end{align}
where $ \psi $ represents the nucleon field, $ \phi $ is the scalar meson field mediating the attractive interaction, $ V_\mu $ is the vector meson field mediating the repulsive interaction, and $ b_\mu $ is the isovector meson field accounting for isospin asymmetry. The nonlinear scalar self-interaction terms involving $ g_2 $ and $ g_3 $ are included to reproduce the empirical properties of nucleonic  matter~\cite{Boguta:1977xi}.
The field strength tensors for the vector mesons are defined as
\begin{align}
V_{\mu \nu} &= \partial_\mu V_\nu - \partial_\nu V_\mu\,, \\
b_{\mu \nu} &= \partial_\mu b_\nu - \partial_\nu b_\mu\,.
\end{align}
To incorporate DM interactions, we extend the Lagrangian by introducing a fermionic DM field $ \chi $ that interacts with the Standard Model via the Higgs portal~\cite{Patt:2006fw}. The extended Lagrangian is given by
\cite{Das:2018frc,Das:2021hnk,Arcadi:2019lka}
\begin{align}
\mathcal{L}_{\text{\rm DM}} =&\ \bar{\chi} \left( i \gamma^\mu \partial_\mu - M_\chi + y h \right) \chi + \frac{1}{2} \partial_\mu h \partial^\mu h - \frac{1}{2} M_h^2 h^2 + f \frac{M_{\rm N}}{v} \bar{\psi} h \psi\,,
\label{eq:lagrangian_dm}
\end{align}
where $ h $ is the Higgs field, $ y $ is the Yukawa coupling between  DM and the  Higgs, $ f $ represents the nucleon-Higgs form factor \footnote{We fix the parameters to $M_{h}=125$~GeV \cite{ParticleDataGroup:2024cfk},  $y=0.07$ and $f=0.35$ as a reasonable choice, see e.g.\ Ref.~\cite{Das:2018frc}.
To assess the sensitivity, varying $ y $ or $ f $ by a factor of 2 modifies the effective DM mass $ M_\chi^* $ by $ \sim 10\% $, leading to small shifts in the mass-radius relation for $ M_\chi = 1000 \, \text{MeV} $. Larger deviations could significantly alter the EoS stiffness, a topic we defer to future work.
}, $ M_{\rm N} $ and $M_{\chi}$ are the nucleon and $\chi$ masses, respectively, and $ v $ is the vacuum expectation value of the Higgs field. 
The Higgs portal parameters ($ M_h = 125 \, \text{GeV} $, $ y = 0.07 $, $ f = 0.35 $) are chosen based on experimental constraints and theoretical consistency. The Higgs mass is set to its measured value \cite{ParticleDataGroup:2024cfk}, while the Yukawa coupling $ y = 0.07 $ and nucleon-Higgs form factor $ f = 0.35 $ are adopted from prior studies (e.g., \cite{Das:2018frc}) as representative values that ensure weak DM-nucleon coupling without violating direct detection bounds. The couplings we have chosen are small enough to avoid the bounds on Higgs decay from collider data. The overall behavior of the equation of state of the mixed  matter has a relatively low sensisivity with respect to the coupling between the two sectors in our set-up. 
The interaction term $ f \frac{M_{\rm N}}{v} \bar{\psi} h \psi $ accounts for the coupling between nucleons and the Higgs field~\cite{Shifman:1978zn}. The RMF model employed here is the NL3. It saturates nuclear matter at $ \rho_0 \approx 0.15 \, \text{fm}^{-3} $ with a binding energy of $ -16.0 \, \text{MeV} $ and supports a maximum neutron star mass of more than  2 $\, M_\odot$, aligning with observational constraints.
While our RMF model with $\omega$, $\rho$, and $\sigma$ meson interactions (Eq.~(\ref{eq:lagrangian_nucleon})) captures the essential physics of nucleonic matter with Higgs portal DM, richer RMF frameworks exist, such as those incorporating additional meson fields or hyperons \cite{Chatterjee:2015pua}. Similarly, alternative dark sector models, e.g., those with different DM interactions \cite{Berryman:2023rmh}, could yield distinct effects. Our choice represents a minimal yet thermodynamically consistent extension to explore DM-nucleon coupling via the Higgs portal, serving as a baseline for future comparisons.

We can now derive the Dirac equation for the nucleons
\begin{equation}
\left[ \gamma^\mu \left( i\partial_\mu - g_v V_0 - g_\rho \tau_3 b_0 \right) - \left( M_{\rm N}^* \right) \right] \psi = 0\,, \label{eq:nucleon_field_eq}
\end{equation}
where the effective mass for nucleons is defined as
\begin{align}
M_{\rm N}^* &= M_{\rm N} + g_s \phi_0 - \frac{f M_{\rm N}}{v} h_0\,. \label{eq:effective_mass_nucleon} 
\end{align}
As for the mesons, the field equations of motion follow from the Euler-Lagrange
equations, where in the mean-field approximation, the meson fields
are replaced by their classical expectation values (denoted by the subscript $0$)\cite{Das:2018frc,Das:2021hnk}
\begin{align}
\phi_0 &= \frac{1}{m_s^2} \left( -g_s \langle \bar{\psi} \psi \rangle - g_2 \phi_0^2 - g_3 \phi_0^3 \right)\,, \label{eq:phi0_solution} \\
V_0 &= \frac{g_v}{m_V^2} \langle \psi^\dagger \psi \rangle = \frac{g_v}{m_V^2} (\rho_p + \rho_n)\,, \nonumber \\
b_0 &= \frac{g_\rho}{m_\rho^2} \langle \psi^\dagger \tau_3 \psi \rangle = \frac{g_\rho}{m_\rho^2} (\rho_p - \rho_n)\,, \nonumber \\
h_0 &= \frac{y \langle \bar{\chi} \chi \rangle + f \frac{M_{\rm N}}{v} \langle \bar{\psi} \psi \rangle}{M_h^2}\,. \nonumber
\end{align}
Here, $ \rho_p $ and $ \rho_n $ are the proton and neutron densities, respectively, and $ \langle \bar{\psi} \psi \rangle $ is the scalar density of nucleons. The values of mass and couplings of all particles in the visible sector are shown in Table~\ref{tab:coup}.

\begin{table}
\centering
\renewcommand{\arraystretch}{1.4}  % Reduced padding for compactness
\small  % Reduced font size
\begin{tabular}{|c|c|c|c|c|c|c|c|c|}
\hline
$M_{\rm N}$ (MeV) & $m_s$ (MeV) & $m_V$ (MeV) & $m_\rho$ (MeV) & $g_s$ & $g_V$ & $g_\rho$ & $g_2$ (fm$^{-1}$) & $g_3$ \\ \hline
939 & 508.2 & 782.5 & 763.0 & 10.22 & 12.87 & 4.47 & -10.43 & -28.89 \\ \hline
\end{tabular}
\caption{Parameters for the nucleonic sector used in Eq.~(\ref{eq:lagrangian_nucleon}), see Refs.~\cite{Das:2018frc,Das:2021hnk}.}
\label{tab:coup}  
\end{table}

The baryon and DM  number densities are given by \cite{Das:2018frc,Das:2021hnk}
\begin{align}
\rho &= \langle \psi^\dagger \psi \rangle = \frac{\gamma}{(2\pi)^3} \int_0^{k_F} d^3k\,, \label{eq:baryon_density} \\
\rho_{\rm N}^{s} &= \langle \bar{\psi} \psi \rangle = \frac{\gamma}{(2\pi)^3} \int_0^{k_F} \frac{M_{\rm N}^*}{\sqrt{M_{\rm N}^{*2} + k^2}} d^3k\,, \label{eq:scalar_density} \\
\rho_{\text{\rm DM}}^s &= \langle \bar{\chi} \chi \rangle = \frac{\gamma}{(2\pi)^3} \int_0^{k_F^{\text{\rm DM}}} \frac{M_\chi^*}{\sqrt{M_\chi^{*2} + k^2}} d^3k\,, \label{eq:dm_scalar_density}
\end{align}
with 
\begin{align}
M_\chi^* &= M_\chi - y h_0\, , \label{eq:effective_mass_dm} 
\end{align}
and where $ \gamma $ ($=2$) is the spin degeneracy factor for fermions, and $ k_F $ and  $ k_{\chi}=k_F^{\text{\rm DM}} $  are the Fermi momenta of nucleons and DM particles, respectively \footnote{ 
The number of nucleons can be obtained from the following integral~\cite{Glendenning:1997wn}:
\begin{eqnarray}
\label{eq:numn}
   N_{\rm N}= \int 4 \pi r^2 \left(1-\frac{2GM(r)}{r}\right)^{-1/2} \rho_{\rm N}(r) \, dr\,.
\end{eqnarray}
Similarly, one can estimate the number of DM particles from~\cite{Glendenning:1997wn}:
\begin{eqnarray}
\label{eq:numdm}
   N_{\rm DM}= \int 4 \pi r^2 \left(1-\frac{2GM(r)}{r}\right)^{-1/2} \rho_{\rm DM}(r) \, dr\,.
\end{eqnarray}
The fraction of DM particles with respect to nucleons is then given by
$f_{\rm DM}={N_{\rm DM}}/{N_{\rm N}}$.
}.
The chemical potentials for nucleons and leptons are given by~\cite{Glendenning:1997wn}
\begin{align}
\mu_p &= g_v V_0 + g_\rho b_0 + \sqrt{k_p^2 + (M_{\rm N}^*)^2}\,, \label{eq:mu_p} \\
\mu_n &= g_v V_0 - g_\rho b_0 + \sqrt{k_n^2 + (M_{\rm N}^*)^2}\,, \label{eq:mu_n} \\
\mu_e &= \sqrt{k_e^2 + m_e^2}\,, \quad \mu_\mu = \sqrt{k_\mu^2 + m_\mu^2}\,. \label{eq:mu_e_mu}
\end{align}

\begin{figure}[htp]
\includegraphics[width=6.50cm]{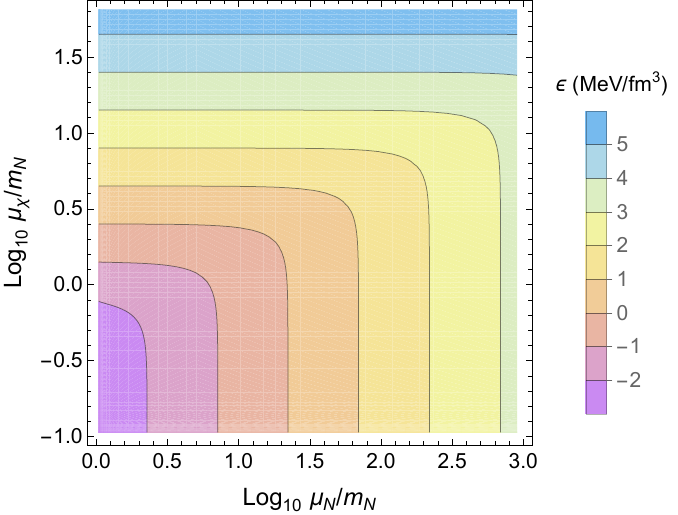}
\includegraphics[width=6.50cm]{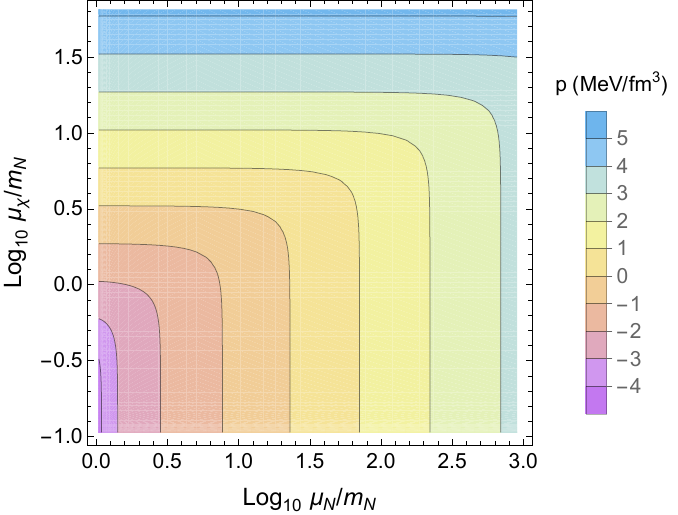}
\includegraphics[width=6.50cm]{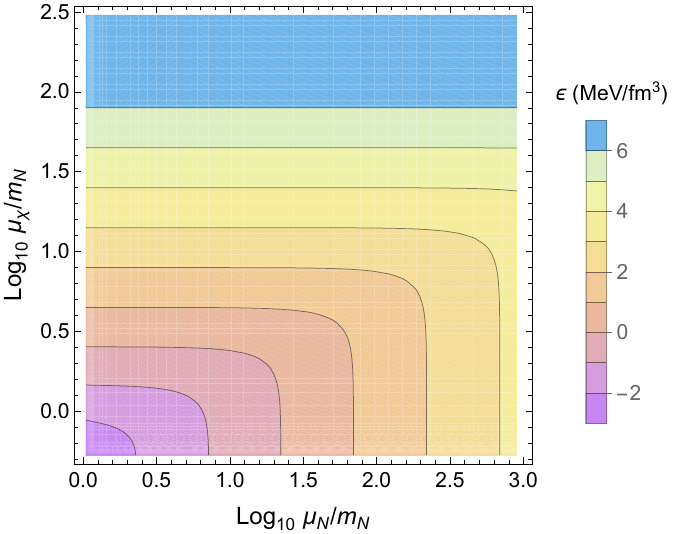}
\includegraphics[width=6.50cm]{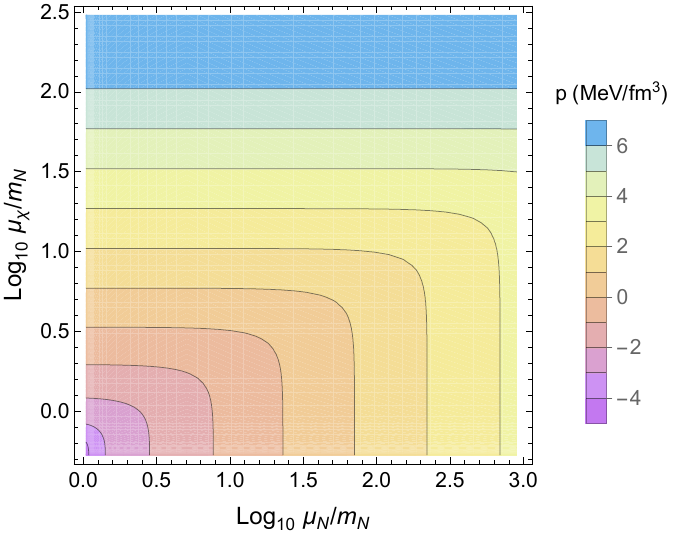}
\includegraphics[width=6.50cm]{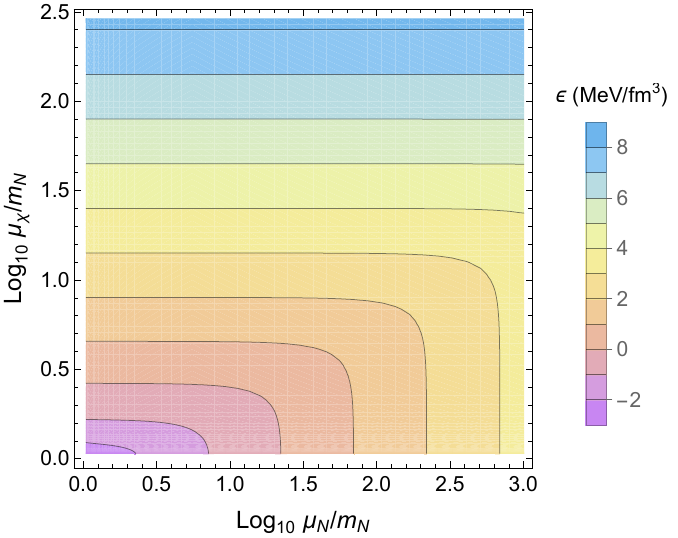}
\includegraphics[width=6.50cm]{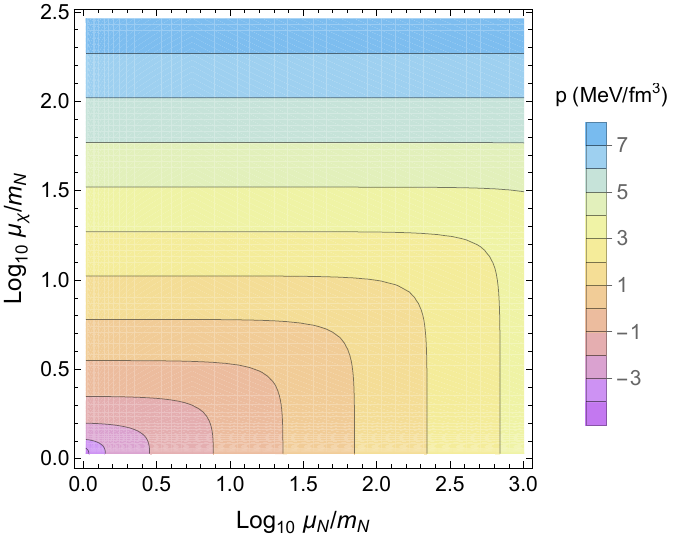}
\includegraphics[width=6.50cm]{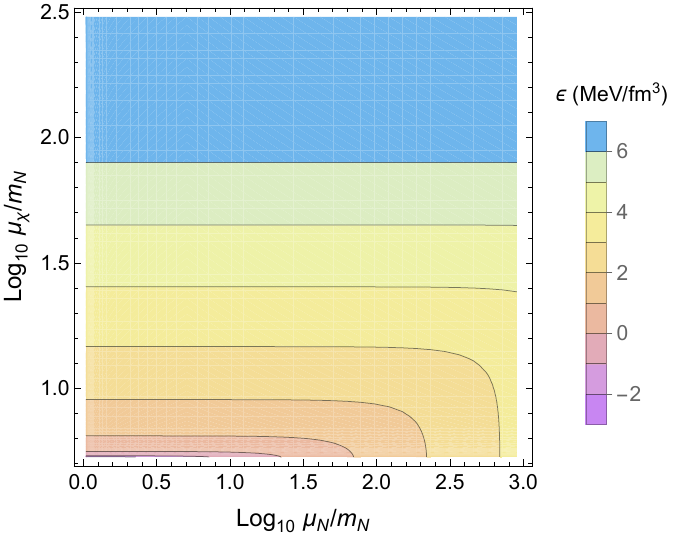}
\includegraphics[width=6.50cm]{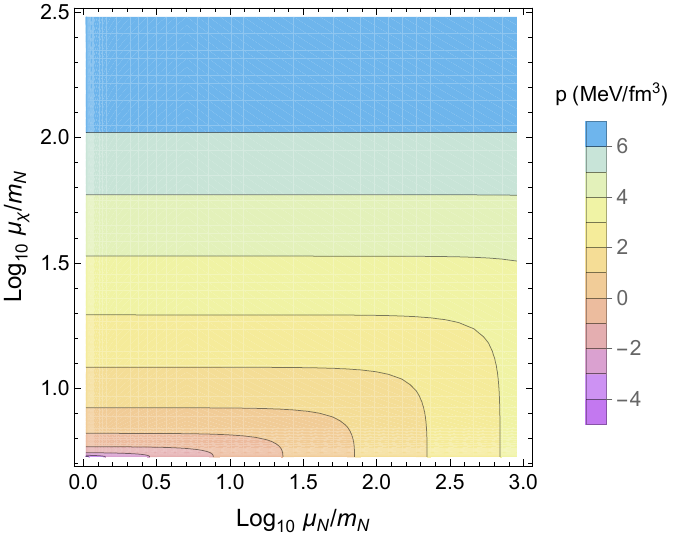}
\centering
\caption{Energy (left panels) and pressure (right panels) densities versus nucleon and DM chemical potentials i.e. $\mu_{\rm N}$ and $\mu_{\chi}$\,. Each row corresponds to a specific DM particle mass $M_{\chi}=100~$MeV (first row), $M_{\chi}=500~$MeV (second row), $M_{\chi}=1000~$MeV (third row) and $M_{\chi}=5000~$MeV (fourth row).  We normalize the DM and nucleon chemical potentials with respect to the nucleon mass. }
\label{fig:eos}
\end{figure}

As for DM, the chemical potential results from
\begin{align}
\mu_\chi &= \sqrt{k_\chi^2 + (M_\chi^*)^2}\,. \label{eq:mu_chi}
\end{align}
With all these ingredients, we can obtain the total energy density $ \epsilon_{\text{tot}} $ and pressure $ p_{\text{tot}} $ and, hence, the EoS. The EoS of the system is calculated by summing contributions from nucleons, DM, mesons,  Higgs, and leptons. Whereas the energy density is given by \cite{Das:2018frc,Das:2021hnk}
\begin{align}
\epsilon_{\text{tot}} =&\ \epsilon_{\text{baryons}} + \epsilon_{\text{\rm DM}} + \epsilon_{\text{mesons+Higgs}} + \epsilon_{\text{leptons}}\,, \label{eq:total_energy_density} \\
\epsilon_{\text{baryons}} =&\ \frac{\gamma}{(2\pi)^3} \int_0^{k_F} d^3k \sqrt{k^2 + (M_{\rm N}^*)^2}\,, \label{eq:epsilon_baryons} \\
\epsilon_{\text{\rm DM}} =&\ \frac{\gamma}{(2\pi)^3} \int_0^{k_F^{\text{\rm DM}}} d^3k \sqrt{k^2 + (M_\chi^*)^2}\,, \label{eq:epsilon_dm} \\
\epsilon_{\text{mesons+Higgs}} =&\ \frac{1}{2} m_s^2 \phi_0^2 + \frac{1}{3} g_2 \phi_0^3 + \frac{1}{4} g_3 \phi_0^4 - \frac{1}{2} m_V^2 V_0^2 - \frac{1}{2} m_\rho^2 b_0^2 + \frac{1}{2} M_h^2 h_0^2\,, \label{eq:epsilon_mesons} \\
\epsilon_{\text{leptons}} =&\ \sum_{l=e,\mu} \frac{1}{\pi^2} \int_0^{k_l} k^2 \sqrt{k^2 + m_l^2} dk\, , \label{eq:epsilon_leptons}
\end{align}
the pressure is obtained from \cite{Das:2018frc,Das:2021hnk}
\begin{align}
p_{\text{tot}} =&\ p_{\text{baryons}} + p_{\text{\rm DM}} + p_{\text{mesons+Higgs}} + p_{\text{leptons}}\,, \label{eq:total_pressure} \\
p_{\text{baryons}} =&\ \frac{\gamma}{3(2\pi)^3} \int_0^{k_F} \frac{k^4 dk}{\sqrt{k^2 + (M_{\rm N}^*)^2}}\,, \label{eq:p_baryons} \\
p_{\text{\rm DM}} =&\ \frac{\gamma}{3(2\pi)^3} \int_0^{k_F^{\text{\rm DM}}} \frac{k^4 dk}{\sqrt{k^2 + (M_\chi^*)^2}}\,, \label{eq:p_dm} \\
p_{\text{mesons+Higgs}} =&\ -\left( \frac{1}{2} m_s^2 \phi_0^2 + \frac{1}{3} g_2 \phi_0^3 + \frac{1}{4} g_3 \phi_0^4 - \frac{1}{2} m_V^2 V_0^2 - \frac{1}{2} m_\rho^2 b_0^2 + \frac{1}{2} M_h^2 h_0^2 \right)\,, \label{eq:p_mesons} \\
p_{\text{leptons}} =&\ \sum_{l=e,\mu} \frac{1}{3\pi^2} \int_0^{k_l} \frac{k^4 dk}{\sqrt{k^2 + m_l^2}}\,. \label{eq:p_leptons}
\end{align}
Note that the negative sign in  $ p_{\text{mesons+Higgs}} $ arises from the meson and Higgs fields contributions to the pressure.

In the core of compact stars, we find $\beta$-equilibrated
charged neutral matter. Consequently, in matter formed by neutrons, protons, electrons and muons the chemical
potentials and the corresponding particle densities satisfy the conditions
\begin{align}
\mu_n &= \mu_p + \mu_e\,, \quad \mu_e = \mu_\mu\,, \\ 
\rho_p &= \rho_e + \rho_\mu\,, \label{eq:charge_neutrality}
\end{align}
where $ \rho_e $ and $ \rho_\mu $ are the electron and muon number densities, respectively.
\label{eq:beta_equilibrium}
Note that our DM candidate is a neutral Majorana fermion, so we do not consider it in $\beta$-equilibrium with ordinary matter.

\section{Results}
\label{sec:results}

\begin{figure}
\includegraphics[width=6.50cm]{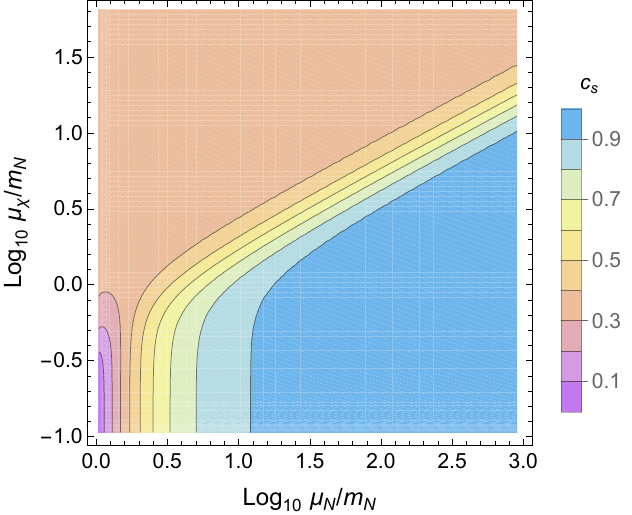}
\includegraphics[width=6.50cm]{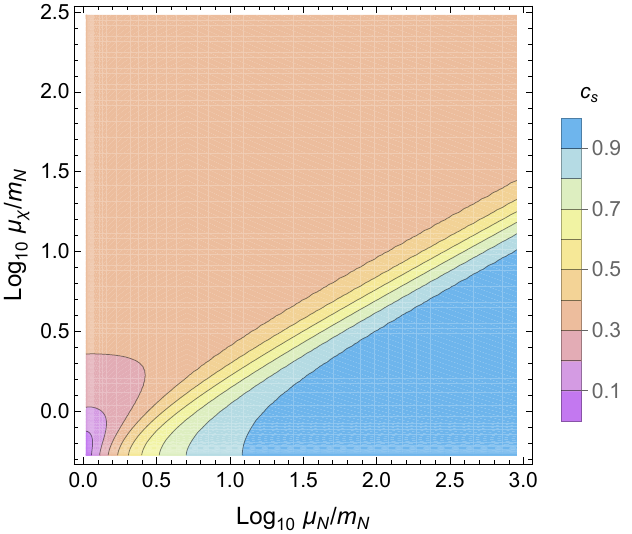}
\includegraphics[width=6.50cm]{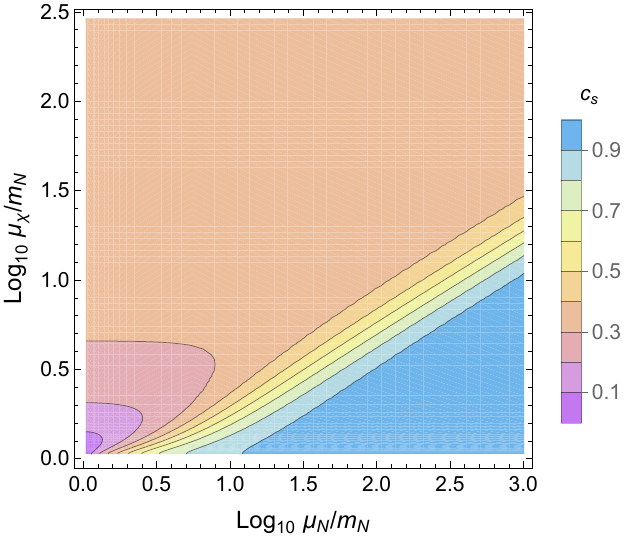}
\includegraphics[width=6.50cm]{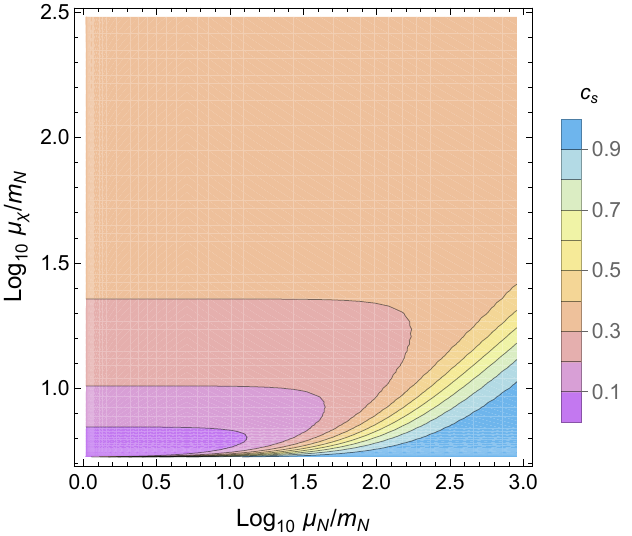}
\centering
\caption{Speed of sound versus nucleon and DM chemical potentials i.e. $\mu_{\rm N}$ and $\mu_{\chi}$\,. Each row is for a specific DM particle mass, that is, $M_{\chi}=100~$MeV (top left), $M_{\chi}=500~$MeV (top right), $M_{\chi}=1000~$MeV (bottom left) and $M_{\chi}=5000~$MeV (bottom right). We normalize the DM and nucleon chemical potentials with respect to the nucleon mass.} 
\label{fig:cs}
\end{figure}

In Fig.~\ref{fig:eos}, we show the EoS for the nucleons and Higgs portal DM model. The results for the energy density and the pressure density with respect to chemical potentials of nucleonic matter and fermionic DM are shown in different rows. Each row corresponds to a specific DM particle mass, that is, $M_{\chi}=100~$MeV (first row), $M_{\chi}=500~$MeV (second row), $M_{\chi}=1000$~MeV (third row) and $M_{\chi}=5000~$MeV (fourth row). 
We use these sets of EoSs in the calculation of the structure equations to find the mass-radius relations and the tidal deformabilities, as we will discuss in Figs.~\ref{fig:MR-tot} and \ref{fig:tidal}. Moreover, we can compute the speed of sound $c_s^2 = dp/d\epsilon$ with respect to $\mu_{\rm N}$ and  $\mu_{\chi}$ for different DM particle masses, as displayed in Fig.~\ref{fig:cs}.

 As shown in Fig.~\ref{fig:cs}, the evolution of $c_s^2$ in an interacting nucleonic and  DM fluids is influenced by the interaction terms in  the EoS. 
For low values of the DM chemical potential $\mu_{\chi}$, the matter is primarily nucleonic. In this regime, the scalar meson field ($\phi$) reduces the effective nucleon mass, leading to a softening of the EoS at low $\mu_{\rm N}$ (see Fig.~\ref{fig:eos}) and resulting in $c_s^2 < \frac{1}{3}$. As $\mu_{\rm N}$ increases, the repulsive vector meson field ($V_\mu$) becomes more dominant, increasing the pressure more than the energy density. This stiffens the EoS and can push $c_s^2$ above $\frac{1}{3}$.
As $\mu_{\chi}$ increases and DM contributions become significant, the coupling of DM with the Higgs field and nucleons modifies the effective DM particle mass ($M_\chi^*$) and alters the EoS. The presence of DM adds to the energy density without a proportional increase in pressure due to its small coupling, effectively softening the EoS, as shown in Fig.~\ref{fig:eos}. Consequently, even at large $\mu_{\chi}$, $c_s^2$ remains below $\frac{1}{3}$, indicating a more compressible medium. 
We also note that the behaviour of $c_s^2$ is similar for all DM particle masses studied, needing bigger values for $\mu_{\rm N}$ to obtain larger $c_s^2$ when the DM particle mass increases.

At this point we should indicate that we only consider the EoS for the core of a neutron star that accumulates DM. Given that the impact of Higgs portal DM on the crust is not well understood,  it is unclear how it can be appropriately included.  Further studies to develop a crust EoS with DM effects  are therefore needed.

\subsection{Radial Profiles and Mass-Radius Relation }
\label{sec:massrad}

\begin{figure}
    \centering
    % Row 1
    \includegraphics[width=7.5cm]{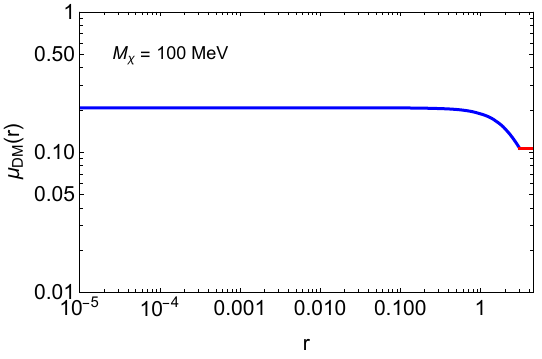}
    \includegraphics[width=7.5cm]{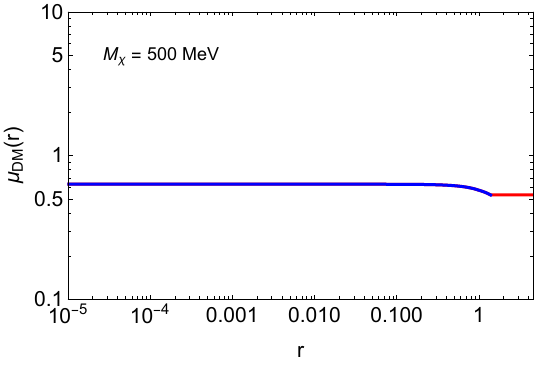}
  
    % Row 2
    \includegraphics[width=7.5cm]{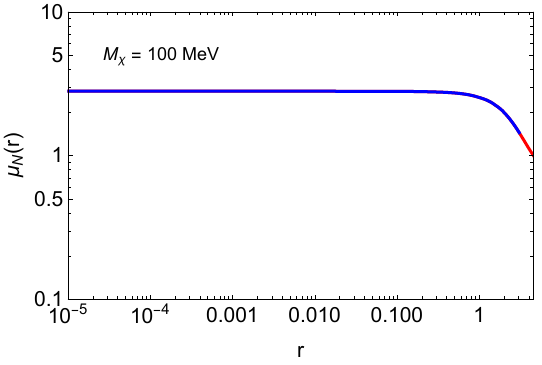}
    \includegraphics[width=7.5cm]{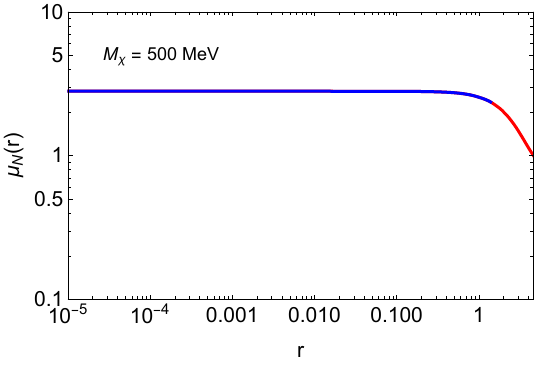}
    
    % % Row 3
    % \includegraphics[width=7.5cm]{nu-r-mdm-100.pdf}
    % \includegraphics[width=7.5cm]{nu-r-mdm-1000.pdf}
    
    % % Row 4
    % \includegraphics[width=7.5cm]{pres-r-mdm-100.pdf}
    % \includegraphics[width=7.5cm]{pres-r-mdm-1000.pdf}
    
    % % Row 5
    % \includegraphics[width=7.5cm]{rho-r-mdm-100.pdf}
    % \includegraphics[width=7.5cm]{rho-r-mdm-1000.pdf}
    
    % Row 6
    \includegraphics[width=7.5cm]{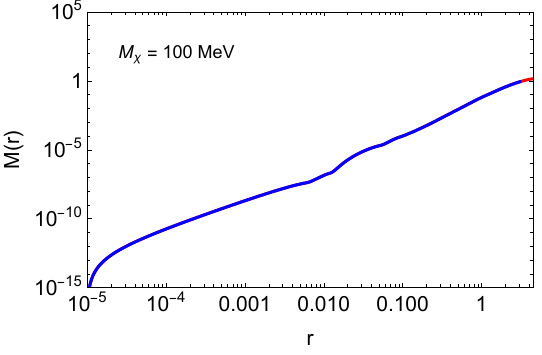}
    \includegraphics[width=7.5cm]{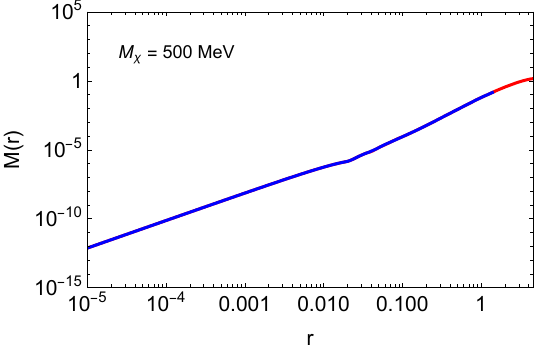}

    % Row 7
    \includegraphics[width=7.5cm]{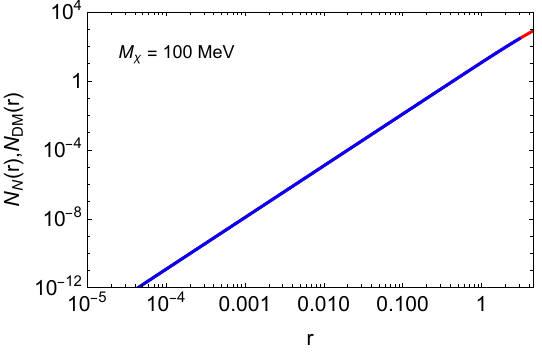}
    \includegraphics[width=7.5cm]{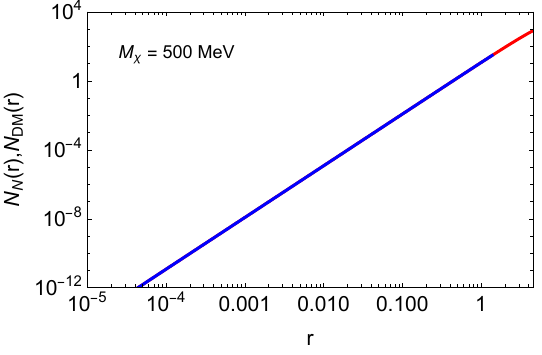}

    % % Row 8
    % \includegraphics[width=7.5cm]{NM-r-mdm-100.pdf}
    % \includegraphics[width=7.5cm]{NM-r-mdm-1000.pdf}

    \caption{A comparison of the radial dependence of various neutron star quantities i.e.  DM and  nucleon chemical potentials, total mass of a neutron star, DM number density and nucleon number density  
 for different DM particle masses. We choose $M_{\chi}=100$~MeV (left panels) and $M_{\chi}=500$~MeV (right panels). The blue parts of the curves indicate the DM and nucleonic matter mixed region and red regions show pure nucleonic matter region. The central chemical potentials we considered here  are $\mu_{\rm N}= M_{\rm N}+1.8$ and  $\mu_{\rm DM}= M_{\chi}+0.1$ (in nucleon mass units).
}
    \label{fig:prof1}
\end{figure}

\begin{figure}
    \centering
    % Row 1
    \includegraphics[width=7.5cm]{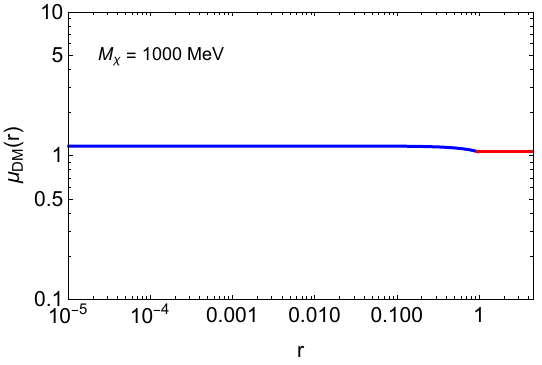}
    \includegraphics[width=7.5cm]{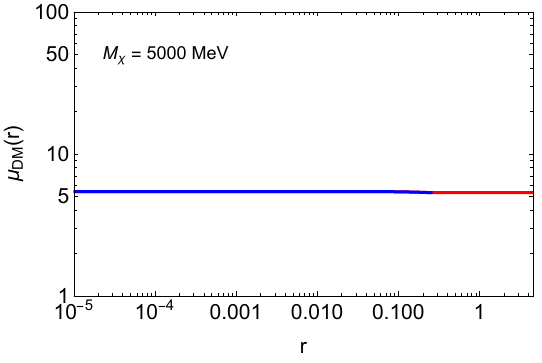}
    \\
    % Row 2
    \includegraphics[width=7.5cm]{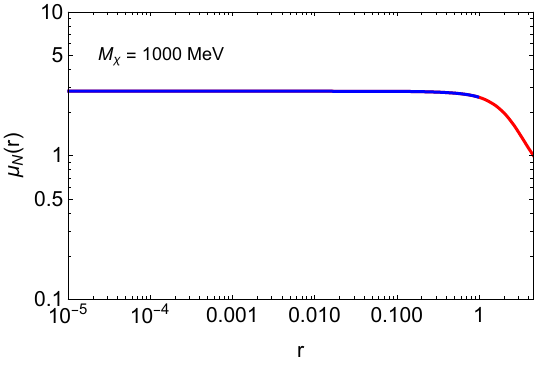}
    \includegraphics[width=7.5cm]{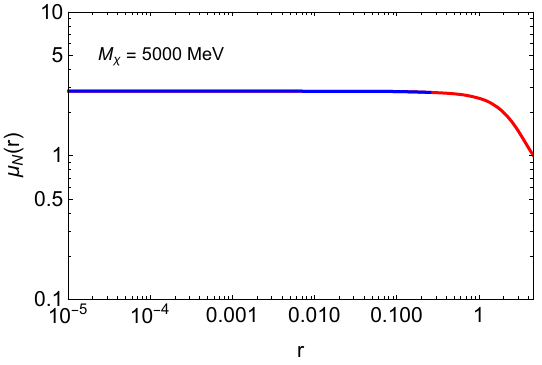}

    % Row 6
    \includegraphics[width=7.5cm]{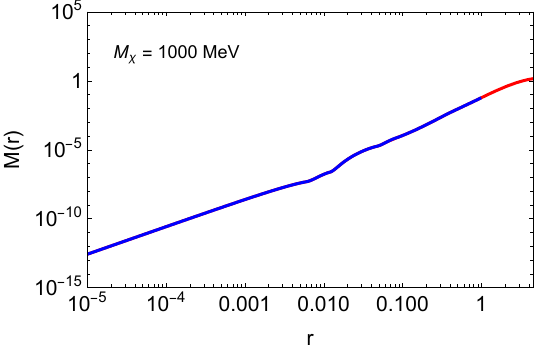}
    \includegraphics[width=7.5cm]{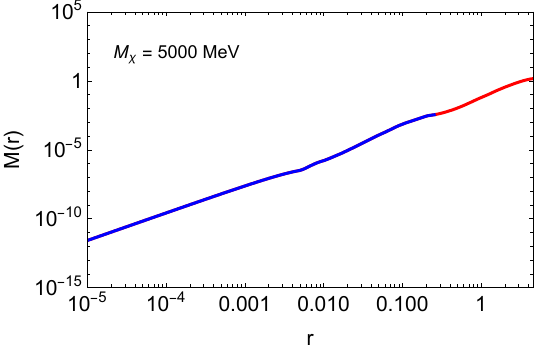}

    % Row 7
    \includegraphics[width=7.5cm]{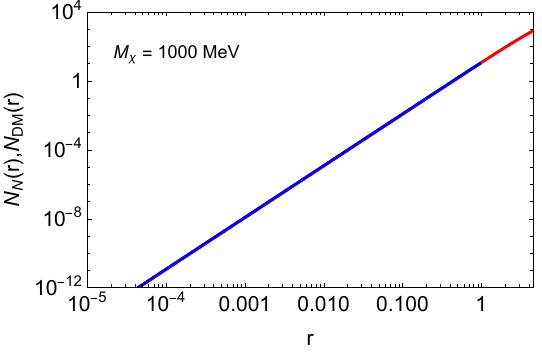}
    \includegraphics[width=7.5cm]{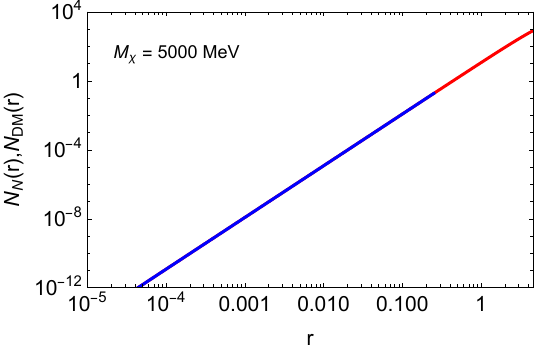}

    \caption{A comparison of the radial dependence of various neutron star quantities i.e. DM and nucleon chemical potentials, total mass of a neutron star, DM number density and nucleonic number density for different DM particle masses. We choose $M_{\chi}=1000$~MeV (left panels) and $M_{\chi}=5000$~MeV (right panels). The blue parts of the curves indicate DM and nucleonic matter mixed region and red parts show pure nucleonic matter region. The central chemical potentials we considered here  are $\mu_{\rm N}= M_{\rm N}+1.8$ and  $\mu_{\rm DM}= M_{\chi}+0.1$ (in nucleon mass units). 
} 
    \label{fig:prof2}
\end{figure}

By solving the structure equations for the chemical potentials, we can extract the radial profiles of the chemical potentials of the dark and visible sectors, as well as the DM and nucleon number densities ($N_{\rm DM}$ and $N_{\rm N}$, respectively) and the total mass of star. These are shown in the different panels of Figs.~\ref{fig:prof1} and \ref{fig:prof2}, where the results of the different columns correspond to DM particles masses of $M_{\chi}=100~$MeV, $M_{\chi}=500~$MeV, 
$M_{\chi}=1000$~MeV and $M_{\chi}=5000~$MeV. Note that the blue lines denote the DM and nucleonic matter mixed region and the red lines indicate
the region where only nucleonic matter is present. These lines result from choosing two initial values for the chemical potentials for DM and nucleons, that is, $\mu_{\rm N}= M_{\rm N}+1.8$ and  $\mu_{\rm DM}= M_{\chi}+0.1$ (in nucleon mass units) \footnote{To calculate the structure equations wherever we need the interpolation of EoS we use a machine learning technique called {\tt Random Forest} algorithm to have a precise estimation of interpolated functions.}.  

\begin{figure}
\includegraphics[width=7.5cm]{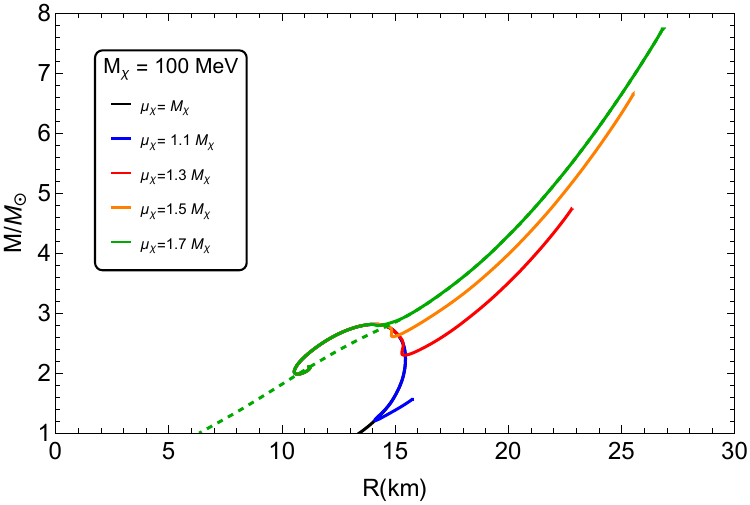}
\includegraphics[width=7.5cm]{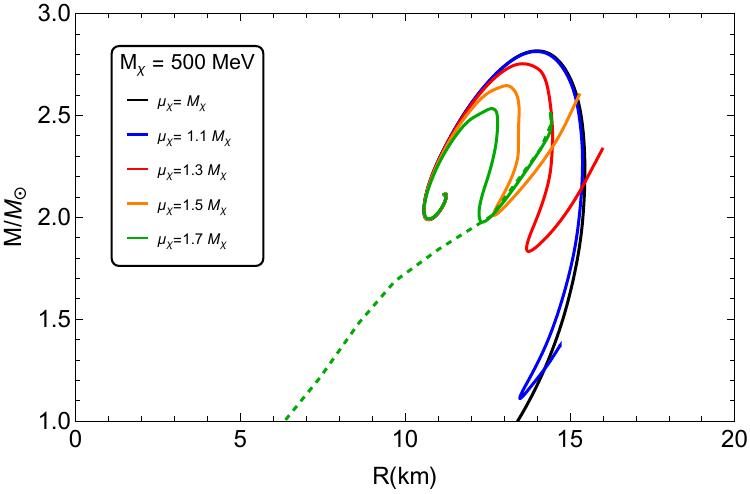}
\includegraphics[width=7.5cm]{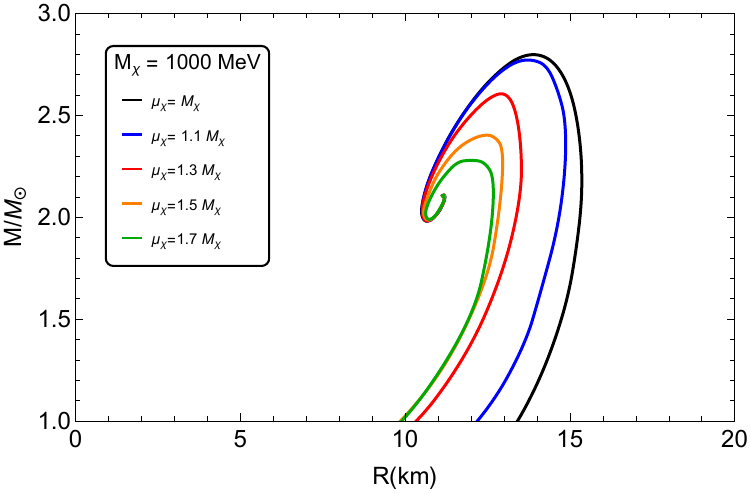}
\includegraphics[width=7.5cm]{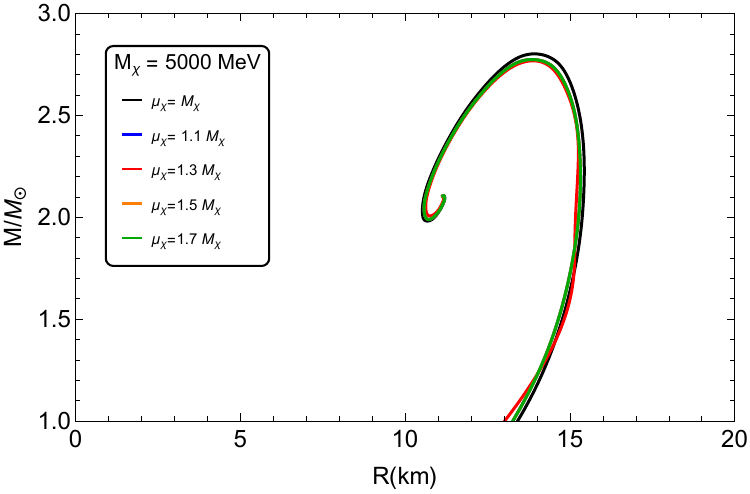}
\centering
\caption{Mass-radius relation for DM-admixed neutron stars. From the top left panel to the bottom right one, we consider for DM particle mass $M_{\chi}=100~$MeV, $M_{\chi}=500~$MeV, $M_{\chi}=1000$~MeV and $M_{\chi}=5000~$MeV.  The different colored lines display the mass-radius relation for various DM chemical potentials at the center (in terms of $M_{\chi}$). Whereas the solid curves are the mass-radius relations for the total mass versus the total radius, the green dashed curves are examples of the total mass versus the radius of DM.  
}
\label{fig:MR-tot}
\end{figure}

As it can be seen in these figures, $\mu_{\chi}$ and $\mu_{\rm N}$ start from the mentioned fixed central values, vary with the distance to the center of the star and become constant when they reach $M_{\chi}$ and $M_{\rm N}$, respectively. In all cases shown, we obtain a DM core surrounded by nucleonic matter. The size of this core depends on the DM particle mass and the relation between the central chemical potentials. In particular, the larger the DM particle mass, the smaller the DM core. Note that if we had explored much larger values for the DM central chemical potential, we would have reached the situation where a DM halo is formed, as we will discuss later. The central chemical potentials ($ \mu_{\rm N} = M_{\rm N} + 1.8 $, $ \mu_{\rm DM} = M_\chi + 0.1 $, in nucleon mass units) are chosen to explore a realistic range for neutron star cores ($ \mu_{\rm N} \sim 1.5-2 \, M_{\rm N} $) and a perturbative DM contribution, reflecting possible DM accumulation scenarios. These values ensure a DM core, while larger $ \mu_{\rm DM} $ leads to halo formation (see Figs. \ref{fig:MR-tot} and \ref{fig:tidal}).
Masses and radii are presented in physical units (solar masses and km) in Figs. \ref{fig:MR-tot} and \ref{fig:tidal}, converted from dimensionless solutions, with scaling independent of $ M_\chi $ to facilitate comparison.

We also show the total mass of the star and the number densities in the third rows of the panels of Figs.~\ref{fig:prof1} and \ref{fig:prof2}. By increasing the DM particle mass from $100$~MeV to $5000$~MeV, a smaller core with DM forms for fixed values of the chemical potentials, since the blue region for DM shrinks and the red region for nucleonic matter expands, as seen in the radial profile of the mass $M(r)$. 
Moreover, the DM number density will be less than the one for nucleons by increasing the DM particle mass as shown in the last row of the panels of same figures as the end of the blue regions moves towards smaller values.
These effects are due to the fact that by increasing the DM particle mass for fixed values of chemical potentials the EoS becomes 
dominated by the nucleonic part of the EoS.

With regard to the mass-radius relation, these are displayed in Fig.~\ref{fig:MR-tot} for different DM particle masses and different central chemical potential ratios (denoted by the different coloured lines in each panel).  The black curves in Fig.~\ref{fig:MR-tot} for all DM particle masses indicate pure nucleonic matter. The other colored curves show admixture of DM with nucleons with different values of nonvanishing chemical potentials. 
 For DM particle masses below the nucleon mass in the top panels of Fig.~\ref{fig:MR-tot}, we observe that the mass-radius relations deviate from the usual neutron star mass-radius configurations as a dark halo appears, exceeding the matter radius. This can be seen in the case of the largest DM chemical potential by comparing the solid green lines for the total mass versus the total radius with respect to the green dashed lines for the total mass versus the DM radius.
For DM particle masses around the nucleon mass a small amount of DM can be sufficient to substantially change the mass-radius relation, as seen in the lower-left panel of Fig.~\ref{fig:MR-tot}. In this case, the increase in the DM central chemical potential leads to more compact DM admixed neutron stars. Also, we note that in this case a dark halo still appears but for masses smaller than a solar mass, that are not represented here. For larger DM particle masses in the lower-right panel of Fig.~\ref{fig:MR-tot}, we do not observe a variation of the mass-radius curves with increasing DM chemical potential, as larger values are needed in order to see any deviation.

We now turn to the fraction of DM with respect to ordinary matter for the maximum mass configurations for radii around $10-15$ km of Fig.~\ref{fig:MR-tot}. These fractions can be computed by means of Eq.~(\ref{eq:numn}) and Eq.~(\ref{eq:numdm}). 
For neutron star matter (black curves), the maximum TOV mass is $ \sim 2.7 \, M_\odot $, consistent with observations. DM admixture softens the EoS, reducing the maximum mass to $ 1.8-2.0 \, M_\odot $ for $ M_\chi = 1000 \, \text{MeV} $ (lower-left panel), still within observational bounds. For lighter DM ($ M_\chi = 100 \, \text{MeV} $), high $ \mu_{\rm DM} $ (green curves) yields masses up to $ 2.5 \, M_\odot $ due to halo formation, potentially exceeding current limits. If one finds a neutron star with mass above 3 or more solar masses it could be explained with the presence of DM. 
The values are shown in Table \ref{tab:fDM}. As it can be seen, increasing the central DM chemical potential (going from blue to green curves)  enhances the fraction of the DM number density with respect to the nucleonic one for DM particle masses smaller than the nucleon one since there can be a halo formation in these cases instead of a core.  
Previous studies of neutron stars with DM have found similar results 
using other DM models when the fraction of DM  with respect to ordinary matter is small and there is a small or no interaction between the two fluids  \cite{Karkevandi:2021ygv,Diedrichs:2023trk,Giangrandi:2022wht,Sagun:2021oml}.

Note that our current model focuses on the core EoS and omits DM effects in the crust due to limited understanding of Higgs portal DM interactions in this regime. While the crust contributes minimally to the total mass and radius for stars with mass  above one solar mass, its inclusion could influence continuous GW signals from non-radial oscillations, as these depend on crust-core interactions. A detailed crust EoS with DM is needed to quantify continuous GW amplitudes, an avenue for future study.

\begin{table}
\centering
\renewcommand{\arraystretch}{1.5} % Adjust the value as needed
\begin{tabular}{|c|c|c|c|c|c|}
\hline
$M_{\chi}$ & $f_{\rm DM}$ (black) & $f_{\rm DM}$ (blue) & $f_{\rm DM}$ (red) & $f_{\rm DM}$ (orange) & $f_{\rm DM}$ (green) \\
\hline
$100~\mathrm{MeV}$ & $0.0$ & $3.11 \times 10^{-2}$ & $1.58 \times 10^{-1}$ & $3.34 \times 10^{-1}$ & $5.55 \times 10^{-1}$ \\
$500~\mathrm{MeV}$ & $0.0$ & $3.26 \times 10^{-2}$ & $1.11 \times 10^{-1}$ & $2.21 \times 10^{-1}$ & $2.44 \times 10^{-1}$ \\
$1000~\mathrm{MeV}$ & $0.0$ & $2.74 \times 10^{-2}$ & $4.78 \times 10^{-2}$ & $3.59 \times 10^{-2}$ & $3.78 \times 10^{-2}$ \\
$5000~\mathrm{MeV}$ & $0.0$ & $6.02 \times 10^{-5}$ & $2.04 \times 10^{-5}$ & $1.08 \times 10^{-5}$ & $7.37 \times 10^{-6}$ \\
\hline
\end{tabular}
\caption{Values of the dark matter number fraction $f_{\rm DM}$ for different DM particle masses $M_{\chi}$ for the maximum mass-radius configurations of Fig.~\ref{fig:MR-tot}. The colors indicate the cases displayed in the figure.}
\label{tab:fDM}
\end{table}

\subsection{Tidal Deformability}
\label{sec:tidal}

\begin{figure}
\includegraphics[width=7.5cm]{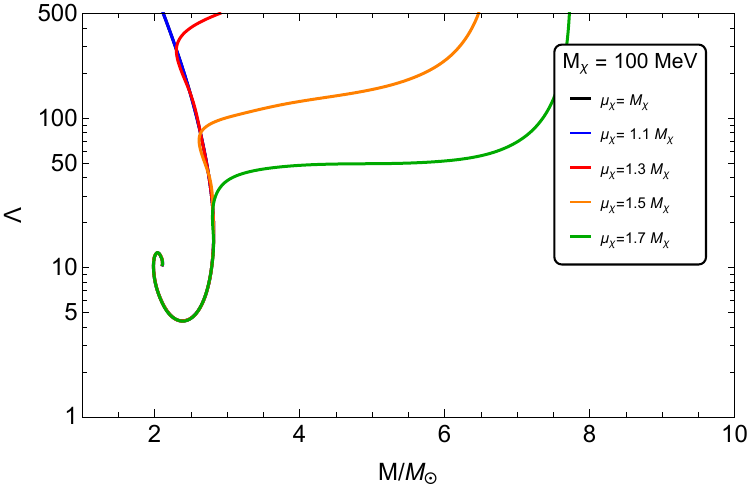}
\includegraphics[width=7.5cm]{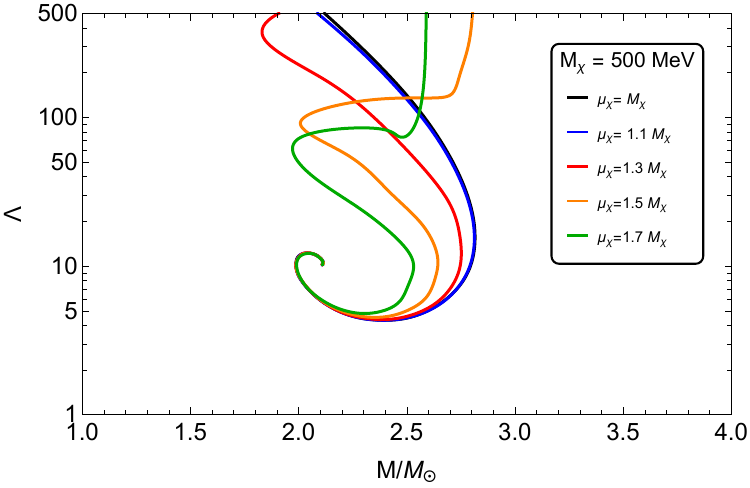}
\includegraphics[width=7.5cm]{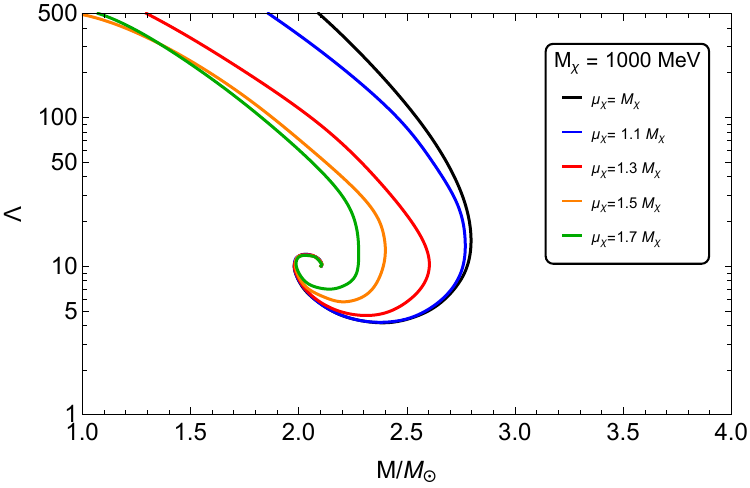}
\includegraphics[width=7.5cm]{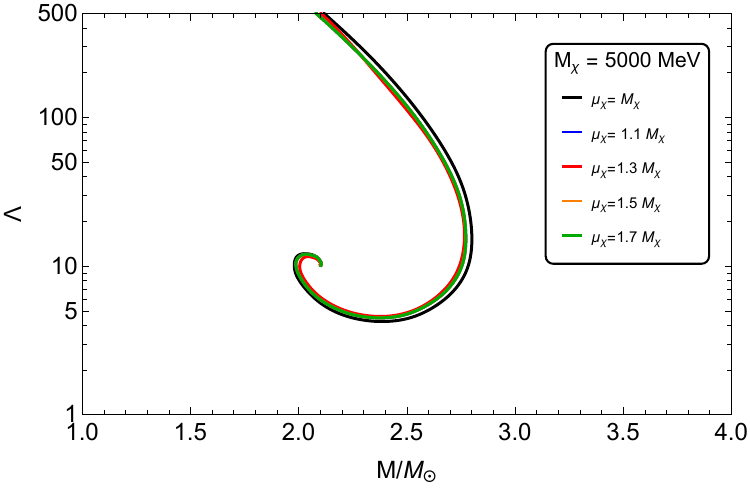}
\centering
\caption{Dimensionless tidal deformability versus the total mass for DM-admixed neutron stars. From the top left panel to the bottom right one, we use for DM particle mass $M_{\chi}=100~$MeV, $M_{\chi}=500~$MeV, $M_{\chi}=1000$~MeV and $M_{\chi}=5000~$MeV. The different colored lines indicate the dimensionless tidal deformability for various DM chemical potentials at the center (in terms of  $M_{\chi})$. }

\label{fig:tidal}
\end{figure}

The tidal deformability $\uplambda$ is given by \cite{Flanagan:2007ix,Hinderer:2007mb,LIGOScientific:2017vwq,Damour:2012yf,Read:2008iy,Lattimer:2006xb,DelPozzo:2013ala,Zacchi:2020dxl} 
\begin{eqnarray}
Q_{ij}=- \uplambda~ \mathcal{E}_{ij}\,,
\end{eqnarray}
where $Q_{ij}$ is the induced quadrupole moment tensor of the compact object, and $\mathcal{E}_{ij}$ is the external tidal field tensor. It quantifies the case with which the compact object (e.g., neutron star) deforms in response to the external tidal field, and it is related to the Love number $k_2$ by \cite{Flanagan:2007ix,Hinderer:2007mb,Zacchi:2020dxl}
\begin{eqnarray}
\uplambda = \frac{2}{3}k_2 R^5\,,
\label{eq:lambdadim}
\end{eqnarray}
where $R$ is the radius of the star. The parameter $\uplambda$ is crucial in gravitational wave astrophysics, especially for binary neutron star mergers, as it influences the late inspiral phase of the gravitational waveform \cite{Flanagan:2007ix,Hinderer:2007mb,LIGOScientific:2017vwq,Damour:2012yf,Read:2008iy}.
To compute the tidal deformability, we have to calculate $ k_2 $. In order to do so, the following differential equation in the dimensionless form has to be solved simultaneously with the structure equations~\cite{Flanagan:2007ix,Hinderer:2007mb,Zacchi:2020dxl}:
\begin{eqnarray}
r'\frac{dy(r')}{dr'} + y(r')^2 + y(r')e^{\lambda(r')}\left\{1 + 4\pi r'^2 \left[p'(r') - \epsilon'(r')\right]\right\} + r'^2 Q(r') &=& 0,
\label{eq:y}
\end{eqnarray}
assuming the initial condition $ y(0) = 2 $. The $ Q $ function that appears in Eq.~(\ref{eq:y}) for $ y $ is defined as:
\begin{eqnarray}
Q(r') &=& 4\pi e^{\lambda(r')}\left[5\epsilon'(r') + 9p'(r') + \frac{\epsilon'(r') + p'(r')}{dp'/d\epsilon'}\right] - 6\frac{e^{\lambda(r')}}{r'^2} - \left[\frac{d\nu(r')}{dr'}\right]^2\, ,
\end{eqnarray}
where $ \lambda(r') $ and $ \nu(r') $ are given by~\cite{Flanagan:2007ix,Hinderer:2007mb,Zacchi:2020dxl}:
\begin{eqnarray}
e^{\lambda(r')} &=& \left[1 - \frac{2M'(r')}{r'}\right]^{-1}, \\ \nonumber
\frac{d\nu}{dr'} &=& \frac{1}{r'}\left[\frac{M'(r') + 4\pi p'(r') r'^3}{r' - 2M'(r')}\right].
\end{eqnarray}
Then,  $k_2$ can be computed from \cite{Flanagan:2007ix,Hinderer:2007mb,Zacchi:2020dxl}
\begin{eqnarray}
k_2 &=& \frac{8C^5}{5} (1 - 2C)^2 [2 + 2C(y - 1) - y] \nonumber \\
&\times& \left\{ 2C[6 - 3y + 3C(5y - 8)] + 4C^3[13 - 11y + C(3y - 2) + 2C^2(1 + y)] \right. \nonumber \\
&& \left. + 3(1 - 2C)^2[2 - y + 2C(y - 1)] \ln(1 - 2C) \right\}^{-1}\,,
\end{eqnarray}
where $y\equiv y(r) |_{r=R}$ and $C$ is the compactness, defined as the ratio of the total mass of the compact star over the radius (in this case, the larger radius of the two species)
\begin{eqnarray}
C=\frac{M(R)}{R}\,.
\end{eqnarray}
The tidal deformability is then given by Eq.~(\ref{eq:lambdadim}), whereas the dimensionless tidal deformability reads
\footnote{The effective tidal deformability parameter  $ \tilde{\Lambda} $ for a binary neutron star merger with separate tidal deformability parameters $\Lambda_1$ and $\Lambda_2$ is given by \cite{LIGOScientific:2017vwq}
\begin{eqnarray}
\tilde{\Lambda} &=& \frac{16 (m_1 + 12m_2)m_1^4 \Lambda_1 + (m_2 + 12m_1)m_2^4 \Lambda_2}{13 (m_1 + m_2)^5}.
\end{eqnarray}}
\begin{eqnarray}
\label{eq:lambda}
\Lambda=\frac{2k_2}{3C^5}\,.
\end{eqnarray}
We show the dimensionless tidal deformability  $\Lambda$ in the four panels of Fig.~\ref{fig:tidal} for different DM particle masses and different central chemical potential ratios (marked by different colored lines in each plot). For DM masses around the nucleon mass, increasing the DM central chemical potential will soften the EoS and lead to smaller masses and radii.  This will then lead to a change in the tidal deformability versus mass, so we obtain smaller masses and tidal deformabilities, as can be read off from Fig.~\ref{fig:tidal}. For DM masses smaller than the nucleon mass, as the central chemical potential of DM increases, the size of the dark halo also expands. At certain chemical potential values, the dark halo can exceed the size of the nucleonic core. This leads to scenarios where the mass-radius relation significantly deviates from that of a typical neutron star, which subsequently impacts the tidal deformability due to the dark halo growing larger than the nucleonic core, as seen in the upper panels of Fig.~\ref{fig:tidal}. For DM masses much larger than the nucleon mass, the tidal deformability is barely unchanged, as expected from the behaviour of the mass-radius relation in Fig.~\ref{fig:MR-tot}.

Compared to the limits extracted from GW170817 measurement of $70 \lesssim \Lambda_{1.4}\lesssim 580$ \cite{LIGOScientific:2018cki}, we find that our results for the dimensionless tidal deformability lie below the upper bound. The overall behaviour of tidal deformability in the presence of DM has been also studied in Refs.~\cite{Das:2018frc,Ellis:2018bkr,Bauswein:2020kor,Karkevandi:2021ygv,Giangrandi:2022wht,Thakur:2024mxs,Rahimi:2024qjl,Ellis:2017jgp,Nelson:2018xtr,Thakur:2023aqm,Thakur:2024ejl,Thakur:2024btu}, but not finding the nonlinear behaviour of the tidal deformability with the total mass which we find for DM particle masses below the nucleon mass.

\section{Summary and Conclusions}
\label{sec:conc}

In this study we have investigated the interplay between nucleonic matter and DM within neutron stars, focusing on the Higgs portal DM model. Our study introduces a thermodynamically consistent approach for solving the structure equations for an interacting system by utilizing the chemical potentials of both nucleons and DM particles, rather than the traditional pressure-based formulation commonly used in the literature. This methodology allows for an accurate and comprehensive analysis of neutron star properties when both visible and DM components weakly interact, while providing predictions that can be tested in future gravitational wave campaigns from binary neutron star mergers.

We have started by revisiting the structure equations, emphasizing the necessity of thermodynamic consistency for systems with interacting nucleonic and DM components by integrating chemical potentials into the structure framework. Using a mean-field model extended through a Higgs portal coupling between nucleons and fermionic DM, we have obtained the EoS of the DM-admixed neutron stars. The DM presence softens the EoS, reducing the pressure and increasing the energy density and decreasing the speed of sound, thus impacting the star’s stability. Solutions to the modified structure equations highlight how DM alters the mass-radius relationship. The behavior depends on the DM particle mass relative to the nucleon mass and its central chemical potential. For certain values of DM particle mass and chemical potential, the DM distribution can form either a core or a halo. When the DM particle mass is larger than the nucleon mass, the structure of the star is not significantly affected. However, for DM particle masses comparable to the nucleon mass, the mass-radius relation shifts, resulting in smaller masses and radii. For DM particle masses much smaller than the nucleon mass but with a large dark chemical potential, a halo may form around the neutron star in addition to its usual structure. These results would have implications for astrophysical observations, as it suggests that the presence of DM within neutron stars could explain the observed compactness that is not accounted for by conventional nucleonic models of neutron stars.

We have also estimated the tidal deformability for neutron stars with varying the DM content, which can be important for gravitational wave analyses, such as the event GW170817 observed by the LIGO and Virgo collaborations \cite{Abbott:2017oio}. An increase in the central chemical potential of DM reduces the tidal deformability, making neutron stars less deformable under tidal forces as long as the dark halo has not formed. For regions where we have a large dark halo like for DM particle masses $100$~MeV and $500$~MeV the total tidal deformability of the star increases. These changes would influence the gravitational wave signal during the inspiral phase of mergers, as studied before \cite{Hinderer:2009ca,Flanagan:2007ix}. The alteration in the tidal deformability values has observational implications, as gravitational wave measurements can constrain the EoS of compact star matter such that an unusual value of the tidal deformability can signal the presence of DM which otherwise could not be reconciled with theoretical nuclear EoS models. Observational constraints on DM halos around neutron stars remain limited. LIGO (GW170817)  and NICER data constrain total mass and radius but do not directly probe DM distributions. Indirectly, pulsar timing or continuous GW signals could reveal DM halo effects via altered rotational dynamics or oscillation modes \cite{Shirke:2023ktu,Shirke:2024ymc}, though no definitive evidence exists yet. For $ M_\chi = 100 \, \text{MeV} $ with large $ \mu_{\rm DM} $, our predicted halos (Fig. \ref{fig:MR-tot}) extend radii beyond typical neutron star sizes ($ R > 15 \, \text{km} $), potentially detectable with future high-precision radius measurements (e.g., from X-ray timing) \cite{Shakeri:2022dwg}.

It is worth noting that our study has been focused on a specific DM model--the Higgs portal fermionic dark matter-- in its mean-field approximation. While this provides valuable insights into the effects of DM on neutron star properties, further attempts are needed to explore other DM candidates and interactions. For instance, considering self-interacting DM, bosonic DM, or asymmetric DM models could reveal additional effects and lead to a more comprehensive understanding of the role of DM in neutron stars.

Additionally, our calculations have not included the effect of DM in the neutron star crust, which could affect the overall structure and observable properties of the star. Incorporating the crust in the EoS for the interacting DM and ordinary matter system and solving the structure equations via the chemical potentials remains an open challenge and serves as an interesting direction for future work. A more detailed treatment of the crust could improve the accuracy of our predictions and enhance the comparison with observational data and future studies on the presence of DM in neutron star. The extension to nonvanishing temperatures, including the study of thermal evolution and cooling of neutron stars containing DM, would also be worth to study. In particular future applications for neutron star merger simulations for interacting fluids of neutron star matter and DM might be an interesting case study as it involves one fluid being kinetically coupled while there are two components with separately conserved number currents. Such a scenario might have unprecedented implications for the generated gravitational wave spectrum.

Finally, our study confirms that DM could affect the structure and observable characteristics of neutron stars. Using our approach may influence  the future study of neutron star rotational dynamics, magnetic field evolution, and emissions of both electromagnetic and gravitational radiation that can be studied further in the future.  In addition, DM can affect the study of pulsar timing and magnetic field of neutron stars that could reveal possible DM interactions.  With advancements in observational techniques, particularly in gravitational wave astronomy, future data from neutron star mergers, precise mass and radius measurements, and studies of magnetic fields and cooling will enable rigorous testing and refinement of theoretical models, ultimately enhancing our understanding of the DM's role in the cosmos.

\acknowledgments

F.H. thanks Raghuveer Garani, Sina M.H. Hajkarim, Davoud Rafiei, Souroush Shakeri and Seokhoon Yun  for useful discussions during  this project.  During the early stages of this project FH was partially supported by the research grant “New Theoretical Tools for Axion Cosmology” under the Supporting TAlent in ReSearch@University of Padova (STARS@UNIPD), Istituto Nazionale di Fisica nucleonice (INFN) through the Theoretical Astroparticle Physics (TAsP) project and the Deutsche Forschungsgemeinschaft (DFG) through the CRC-TR 211 project number 315477589-TRR 211. 
F.H. is supported by Homer Dodge postdoctoral fellowship and in part by DOE grant DE-SC0009956. 
F.H. also thanks the organizers of workshop of Center for Theoretical Underground Physics and Related Areas (CETUP* - 2024), The Institute for Underground Science at Sanford
Underground Research Facility (SURF), Lead, South Dakota 
for their hospitality and financial support. Moreover, he thanks the organizers of the Mitchell
Conference in May 2024 at Texas A \& M University for their hospitality and support during
the initial stages of this project. 
L.T. acknowledges support from CEX2020-001058-M (Unidad de Excelencia ``Mar\'{\i}a de Maeztu") and PID2022-139427NB-I00 financed by the Spanish MCIN/AEI/10.13039/501100011033/FEDER,UE as well as from the Generalitat de Catalunya under contract 2021 SGR 171 and  from the Generalitat Valenciana under contract CIPROM/2023/59.
L.T. and J.S.B. acknowledge support  by the CRC-TR 211 'Strong-interaction matter under extreme conditions'- project Nr. 315477589 - TRR 211.

\bibliography{biblio-jcap.bib}

\end{document}